\documentclass[12pt,fleqn]{article}
\usepackage{amsmath}
\usepackage{amsfonts}
\usepackage{amsthm}
\usepackage{verbatim}
\usepackage{epsfig}
\usepackage{natbib}

\makeatletter
\def\monthname{\ifcase\month\or
January\or February\or March\or April\or May\or June\or
July\or August\or September\or October\or November\or December\fi}
\makeatother

\makeatletter
\def\@sect#1#2#3#4#5#6[#7]#8{\ifnum #2>\c@secnumdepth
     \let\@svsec\@empty\else
     \refstepcounter{#1}\edef\@svsec{\csname the#1\endcsname. \hskip 0.4em}\fi
     \@tempskipa #5\relax
      \ifdim \@tempskipa>\z@
        \begingroup #6\relax
          \@hangfrom{\hskip #3\relax\@svsec}{\interlinepenalty \@M #8\par}%
        \endgroup
       \csname #1mark\endcsname{#7}\addcontentsline
         {toc}{#1}{\ifnum #2>\c@secnumdepth \else
                      \protect\numberline{\csname the#1\endcsname}\fi
                    #7}\else
        \def\@svsechd{#6\hskip #3\relax  
                   \@svsec #8\csname #1mark\endcsname
                      {#7}\addcontentsline
                           {toc}{#1}{\ifnum #2>\c@secnumdepth \else
                             \protect\numberline{\csname the#1\endcsname}\fi
                       #7}}\fi
     \@xsect{#5}}
\makeatother

\makeatletter
\renewcommand{\section}{\@startsection{section}{1}{0mm}{-\baselineskip}{0.25\baselineskip}{\raggedright\normalfont\normalsize\bf}}
\renewcommand{\subsection}{\@startsection{subsection}{2}{0mm}{-\baselineskip}{0.05\baselineskip}{\raggedright\normalfont\normalsize\itshape}}
\renewcommand{\subsubsection}{\@startsection{subsubsection}{3}{0mm}{-\baselineskip}{0.05\baselineskip}{\raggedright\normalfont\small\bf}}
\def\@begintheorem#1#2{\trivlist \item[\hskip \labelsep{\bf #1\ #2:}]\it}
\makeatother

\makeatletter
\def\blfootnote{\xdef\@thefnmark{}\@footnotetext}
\makeatother

\renewenvironment{abstract}
 {\begin{center}\normalsize\bf\text{Abstract}
 \end{center}\begin{quote}\normalsize}
 {\end{quote}}

\begin{document}
\vspace*{0.2cm}

\setcounter{page}{0}
\thispagestyle{empty}
\vskip 10pt
\centerline{\large\bf On Statistical Non-Significance}
  \begin{center}%
    {\large

\par}%
    \blfootnote{\hspace*{-0.25in}Alberto Abadie, Department of Economics, MIT, abadie@mit.edu. We thank Joshua Angrist, Gary Chamberlain, Amy Finkelstein, Guido Imbens, and Ben Olken for comments and discussions. A version of this article aimed at an economics readership has circulated under the title ``Statistical Non-Significant in Empirical Economics''.
}
    {\large
     \lineskip .75em%
      \begin{tabular}[t]{c}%
       Alberto Abadie\\
       MIT\\
      \end{tabular}

      \par}%
     \vskip 1em%
      {\large \monthname \ \number\year} \par
       \vskip 2em%
  \end{center}\par

\begin{abstract}
Significance tests are probably the most extended form of inference in empirical research, and significance is often interpreted as providing greater informational content than non-significance. In this article we show, however, that rejection of a point null often carries very little information, while failure to reject may be highly informative. This is particularly true in empirical contexts  where data sets are large and where there are rarely reasons to put substantial prior probability on a point null. Our results challenge the usual practice of conferring point null rejections a higher level of scientific significance than non-rejections. In consequence, we advocate a visible reporting and discussion of non-significant results in empirical practice. 
\end{abstract}

\setcounter{page}{1}
\addtolength{\baselineskip}{0.5\baselineskip}

\section{Introduction}
\label{section:introduction}

Non-significant empirical results (usually in the form of  $t$-statistics smaller than 1.96) relative to some null hypotheses of interest (usually zero coefficients) are notoriously hard to publish in professional/scientific journals \citep[see, e.g.,][]{ziliak2008cult}. This state of affairs is in part maintained by the widespread notion that non-significant results are non-informative. After all, lack of statistical significance derives from the absence of extreme or surprising outcomes under the null hypothesis. In this article, we argue that this view of statistical inference is misguided. In particular, we show that non-significant results are informative, and argue that they are more informative than significant results in scenarios that are common, even prevalent, in empirical practice.  

To discuss the informational content of different statistical procedures, we formally adopt a limited information Bayes perspective. In this setting, agents representing journal readership or the scientific community have priors, $\mathcal P$, over some parameters of interests, $\theta\in\Theta$. That is, a member $p$ of $\mathcal P$ is a probability density function (with respect to some appropriate measure) on $\mathcal P$. While agents are Bayesian, we will consider a setting where journals report frequentist results, in particular, statistical significance. Agents construct limited information Bayes posteriors based on the reported results of significance tests. We will deem a statistical result informative when it has the potential to substantially change the prior of the agents over a large range of values for $\theta$. 

Notice, that, like \cite{ioannidis2005why} and others, we restrict our attention to the effect of statistical significance on beliefs. We adopt this framework not because we believe it is (always) representative of empirical practice (in fact, journals typically report additional statistics, beyond statistical significance), but because isolating the informational content of statistical significance has immediate implications for how we should interpret its occurrence or lack of it. Correct interpretation of statistical significance is important because, while many other statistics are reported in practice, the scientific discussion of empirical results is often framed in terms of statistical significance of some parameters of interest and non-significant results may be under-reported as discussed above. 

\section{A Simple Example}\label{section:normal}
\begin{figure}
\centering
  \includegraphics[width=4.6in,keepaspectratio=1 ]{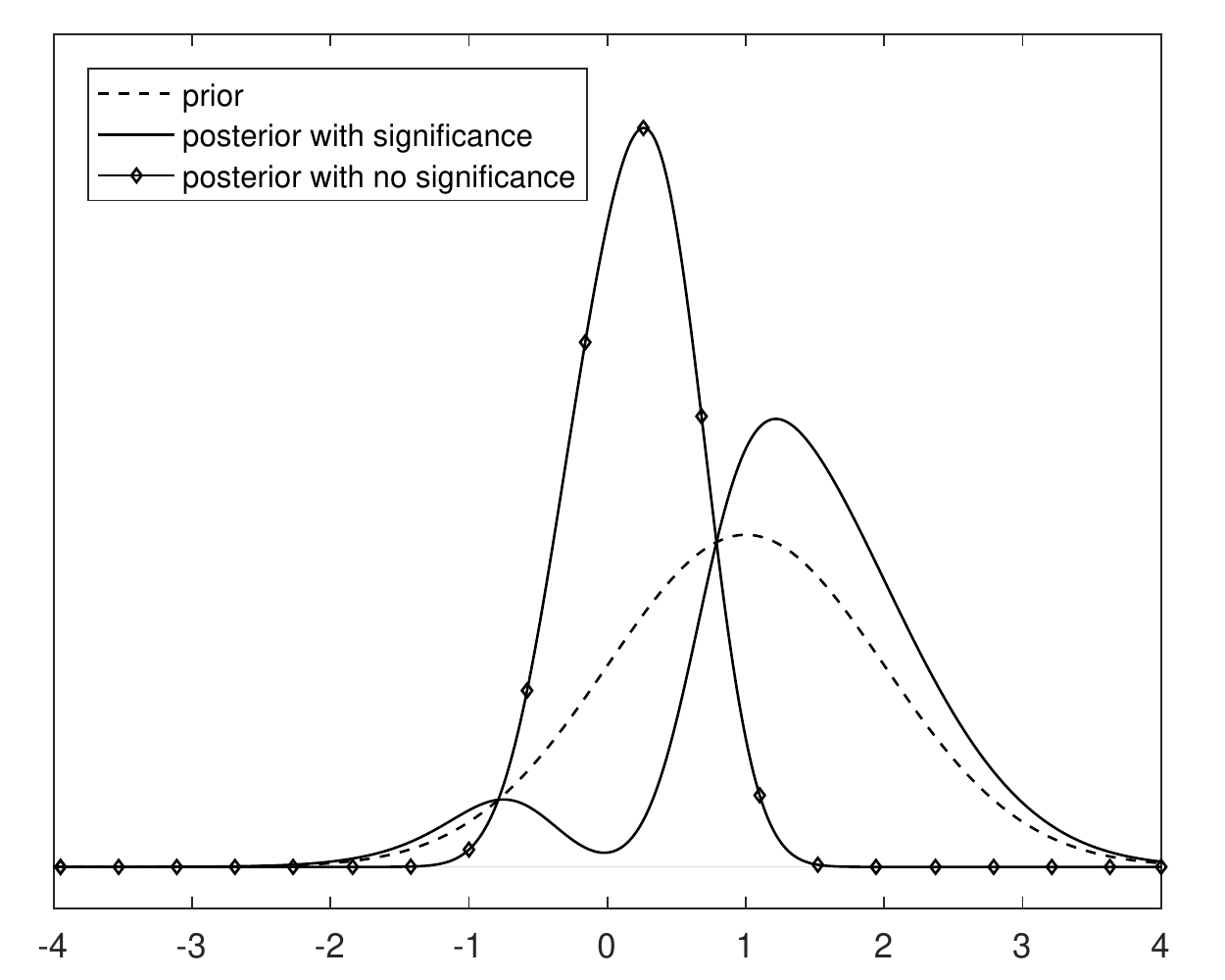}
\caption{Posterior Distributions After a Significance Test}
\label{fig:posterior}
\end{figure}

In this section, we consider a simple example with Normal priors and data that captures the essence of our argument. In section \ref{section:General_Case} we will consider the case where the priors and the distribution of the data are not restricted to be in a particular parametric family. Assume an agent has a prior $\theta\sim N(\mu,\sigma^2)$ on $\theta$, with $\sigma^2>0$. A researcher observes $n$ independent measurement of $\theta$ with Normal errors mutually independent and independent of $\theta$, and with variance normalized to one. That is, $x_1,\ldots, x_n$ are independent $N(\theta,1)$. Let
\[
\widehat\theta=\frac{1}{n}\sum_{i=1}^n x_i\sim N(\theta,1/n).
\]
$\theta$ is deemed significant if $\sqrt{n}|\widehat\theta|> c$, for some $c>0$. In empirical practice, $c$ is often equal to 1.96, the $0.975$-quantile of the Standard Normal distribution. Suppose a journal reports on statistical significance. We will calculate the limited information posteriors of the agents conditional on significance and lack thereof. These posteriors are the distributions of $\theta$ conditional on $\sqrt{n}|\widehat\theta|> c$ and $\sqrt{n}|\widehat\theta|\leq c$. First, notice that
\begin{align*}
\Pr(\sqrt{n}|\widehat\theta|> c|\theta)&=\Pr(\widehat\theta> c/\sqrt{n}|\theta)+\Pr(-\widehat\theta> c/\sqrt{n}|\theta)\\
&=\Phi(\sqrt{n}\theta-c)+\Phi(-\sqrt{n}\theta-c).
\end{align*}
Therefore,\footnote{This calculation uses the following fact of integration
\[
\int \Phi\left(\frac{\lambda-\theta}{\xi}\right)\frac{1}{\sigma}\phi\left(\frac{\theta-\mu}{\sigma}\right)d\theta
=\Phi\left(\frac{\lambda-\mu}{\sqrt{\sigma^2+\xi^2}}\right)
\]
for arbitrary real $\lambda$ and $\mu$ and positive $\sigma$ and $\xi$. Alternatively, the result can be easily derived after noticing that the distribution of $\widehat\theta$ integrated over the prior is Normal with mean $\mu$ and variance $\sigma^2+1/n$.}
\begin{equation}
\label{equation:ProbRej}
\Pr(\sqrt{n}|\widehat\theta|> c)=\Phi\Bigg(\frac{\sqrt{n}\mu-c}{\sqrt{1+n\sigma^2}}\Bigg)+\Phi\Bigg(\frac{-\sqrt{n}\mu-c}{\sqrt{1+n\sigma^2}}\Bigg).
\end{equation}
The limited information posteriors given significance and non-significance are:
\[
p\big(\theta\big|\sqrt{n}|\widehat\theta|>c\big)
=\frac{\displaystyle\frac{1}{\sigma}\phi\Bigg(\displaystyle\frac{\theta-\mu}{\sigma}\Bigg)\Big(\Phi(\sqrt{n}\theta-c)+\Phi(-\sqrt{n}\theta-c)\Big)}
{\Phi\Bigg(\displaystyle\frac{\sqrt{n}\mu-c}{\sqrt{1+n\sigma^2}}\Bigg)+\Phi\Bigg(\displaystyle\frac{-\sqrt{n}\mu-c}{\sqrt{1+n\sigma^2}}\Bigg)},
\]
and
\[
p\big(\theta\big|\sqrt{n}|\widehat\theta|\leq c\big)
=\frac{\displaystyle\frac{1}{\sigma}\phi\Bigg(\displaystyle\frac{\theta-\mu}{\sigma}\Bigg)\Big(1-\Phi(\sqrt{n}\theta-c)-\Phi(-\sqrt{n}\theta-c)\Big)}
{1-\Phi\Bigg(\displaystyle\frac{\sqrt{n}\mu-c}{\sqrt{1+n\sigma^2}}\Bigg)-\Phi\Bigg(\displaystyle\frac{-\sqrt{n}\mu-c}{\sqrt{1+n\sigma^2}}\Bigg)}.
\]
The two posteriors, along with the Normal prior, are plotted in Figure 
\ref{fig:posterior} for $\mu=1$, $\sigma=1$, $c=1.96$, and $n=10$. This figure illustrates the informational value of a significance test. Rejection of the null carves probability mass around zero in the limited information posterior, while failure to reject concentrates probability mass around zero. Notice that failure to reject carries substantial information, even in the rather under-powered setting generated by the values of $\mu$, $\sigma$, $c$, and $n$ adopted for Figure \ref{fig:posterior}, which imply $\Pr(\sqrt n|\widehat\theta|> c\big)=0.7028$.

\begin{figure}
\centering
  \includegraphics[width=4.6in,keepaspectratio=1 ]{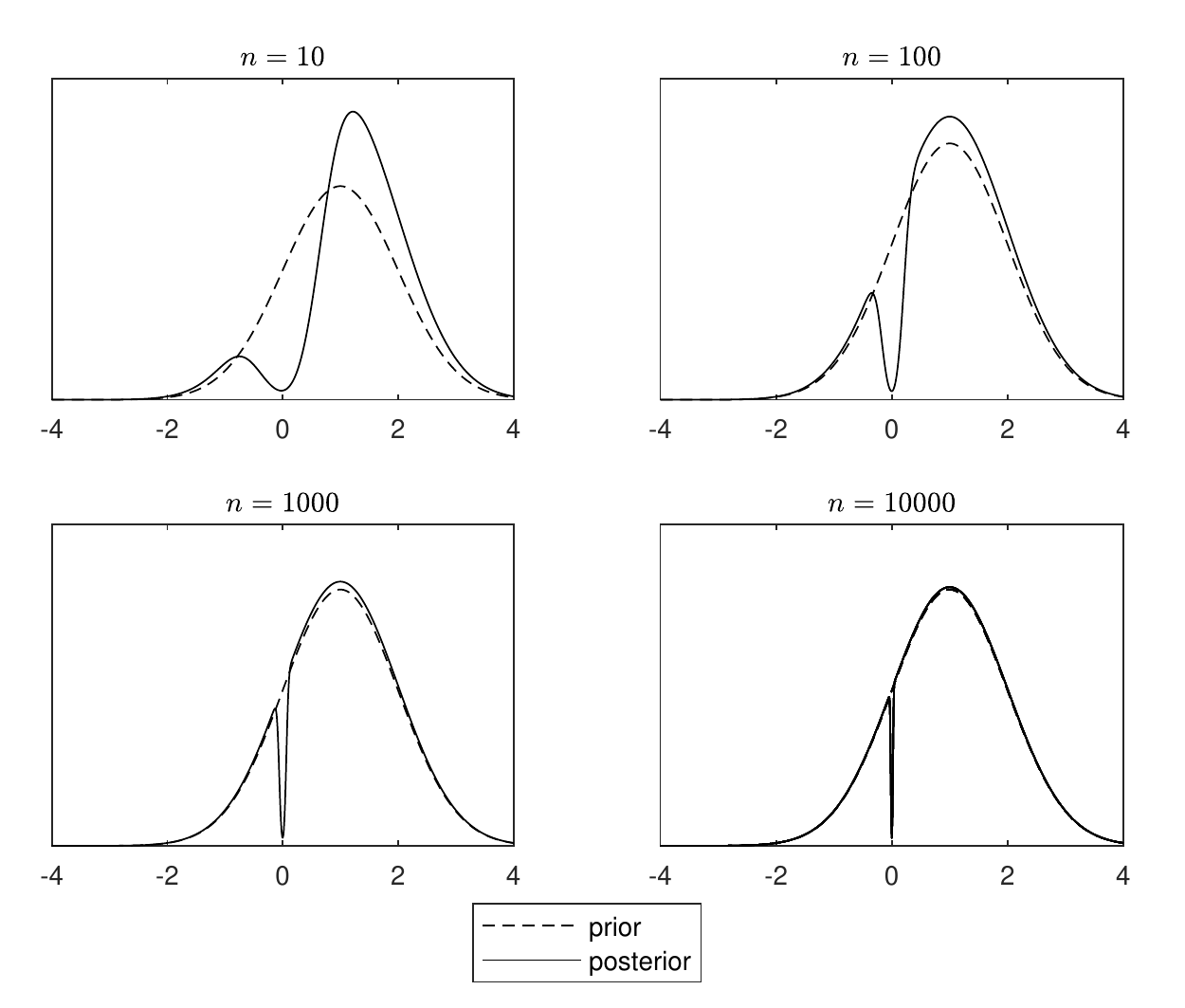}
\caption{Prior and Posterior with Significance for Different Sample Sizes}
\label{fig:postN}
\end{figure}

Figure \ref{fig:postN} shows how prior and posteriors after significance compare as a function of the sample size. When $n$ is small, significance affects the posterior over a large range of values. When $n$ is large, significance provides only local to zero information. That is, significance is not informative in large samples. This is explained by the fact that the probability of rejection in equation (\ref{equation:ProbRej}) converges to one as the sample size increases. By the law of total probability, it follows that conditional on non-significance probability mass concentrates around zero as $n$ increases. That is, the occurrence of an event that is very unlikely given the prior has a large effect on beliefs. 

The full information posterior is
\[
p\big(\theta|x_1,\ldots, x_n\big)= \frac{1}{\sigma_n}\phi\Bigg(\displaystyle\frac{\theta-\mu_n}{\sigma_n}\Bigg),
\]
where
\[
\mu_n
=\frac{\mu+n\sigma^2\widehat\theta}{1+n\sigma^2},
\]
and
\[
\sigma^2_n=\frac{\sigma^2}{1+n\sigma^2}.
\]
So, in this very particular context, knowledge of the $t$-ratio ($\sqrt{n}\widehat\theta$) is sufficient to go back to the full information posterior. The same is true for the combined information given by the $P$-value, $2\Phi(-\sqrt n |\widehat\theta|)$, and the sign of $\widehat\theta$. 

These results have immediate counterparts in in large samples settings with asymptotically Normal distributions. They can also be generalized to non-parametric settings, as we demonstrate in the next section. 

\section{General Case}
\label{section:General_Case}
To extend the results of the previous section beyond Normal priors and data, we will consider a test statistic, $\widehat T_n$, such that
\begin{align*}
\Pr\big(\widehat T_n > c\big| \theta=0\big)&\rightarrow \alpha,\\
\intertext{and}
\Pr\big(\widehat T_n > c\big| \theta, \theta\neq 0\big)&\rightarrow 1.
\end{align*}
That is, we consider significance tests that are consistent under fixed alternatives and have asymptotic size equal to $\alpha$. Let $p(\cdot)$ be a prior on $\theta$, and $p(\cdot | \widehat T_n > c)$ and $p(\cdot | \widehat T_n \leq c)$ be the limited information posteriors under significance and non-significance, respectively. 

\subsection{Continuous Prior}
We will first assume a prior that is absolutely continuous with respect to the Lebesgue measure, with a version of the density that is positive and continuous at zero. By dominated convergence, we obtain:
\[
\Pr\big(\widehat T_n > c\big)\rightarrow 1.
\]
We first derive the posterior densities under significance,
\begin{align*}
p(0|\widehat T_n > c)&=\frac{\Pr\big(\widehat T_n > c\big| \theta=0\big)}{\Pr\big(\widehat T_n > c\big)}p(0)\rightarrow \alpha\, p(0),\\
\intertext{and }
p(\theta|\widehat T_n > c)&=\frac{\Pr\big(\widehat T_n > c\big| \theta\big)}{\Pr\big(\widehat T_n > c\big)}p(\theta)\rightarrow p(\theta),
\end{align*}
for $\theta\neq 0$.
So, again, significance only changes beliefs locally around zero. The posterior densities after non-significance are
\begin{align*}
p(0|\widehat T_n \leq c)&=\frac{\Pr\big(\widehat T_n \leq c\big| \theta=0\big)}{\Pr\big(\widehat T_n \leq c\big)}p(0)\rightarrow \infty,\\
\intertext{and}
p(\theta|\widehat T_n \leq c)&=\frac{\Pr\big(\widehat T_n \leq c\big| \theta\big)}{\Pr\big(\widehat T_n \leq c\big)}p(\theta)
\end{align*}
for $\theta\neq 0$. Typically, for $\theta\neq 0$ (using large deviation results)
\[
-\frac{1}{n} \log\left(\Pr\big(\widehat T_n \leq c\big| \theta\big)\right)\rightarrow d_\theta,
\]
with $0<d_\theta<\infty$. Therefore, $\Pr\big(\widehat T_n \leq c\big| \theta\big)$ converges to zero exponentially for $\theta\neq 0$. Let 
\[
\beta_n(\theta) = \Pr(\widehat T_n\leq c|\theta)
\]
be the probability of Type II error (one minus the power). Assume that
\[
\int \liminf\limits_{n\rightarrow\infty}\beta_n(z/\sqrt n)\, dz>0.
\]
This rules out perfect local asymptotic power. Then, by change of variable $z=n^{1/2}\theta$ and Fatou's lemma, we obtain\footnote{For the second to last equality, notice that if $a_n\geq 0$ and $b_n\rightarrow b>0$ as $n\rightarrow \infty$, then
\[
\liminf\limits_{n\rightarrow\infty} (a_n b_n) =  \liminf\limits_{n\rightarrow\infty} a_n \lim\limits_{n\rightarrow\infty} b_n.
\]
}
\begin{align*}
\liminf\limits_{n\rightarrow\infty}n^{1/2}\Pr(\widehat T_n\leq c) &= \liminf\limits_{n\rightarrow\infty} n^{1/2}\int \beta_n(\theta)\, p(\theta)\, d\theta\\\
&=\liminf\limits_{n\rightarrow\infty}\int \beta_n(z/\sqrt n)\, p(z/\sqrt n)\, dz\\
&\geq \int \liminf\limits_{n\rightarrow\infty}(\beta_n(z/\sqrt n)\, p(z/\sqrt n))\, dz\\
&= \int \liminf\limits_{n\rightarrow\infty}\beta_n(z/\sqrt n)\, \lim\limits_{n\rightarrow \infty}p(z/\sqrt n)\, dz\\
&=p(0) \int \liminf\limits_{n\rightarrow\infty}\beta_n(z/\sqrt n)\,  dz >0.
\end{align*}
It follows that 
\[
p(\theta|\widehat T_n \leq c)\rightarrow 0,
\]
for $\theta\neq 0$.

That is, like in the Normal case of section \ref{section:normal}, conditional on non-significance the posterior converges to a degenerate distribution at zero. 

\begin{figure}[h!]
\centering
  \includegraphics[width=4.8in,keepaspectratio=1 ]{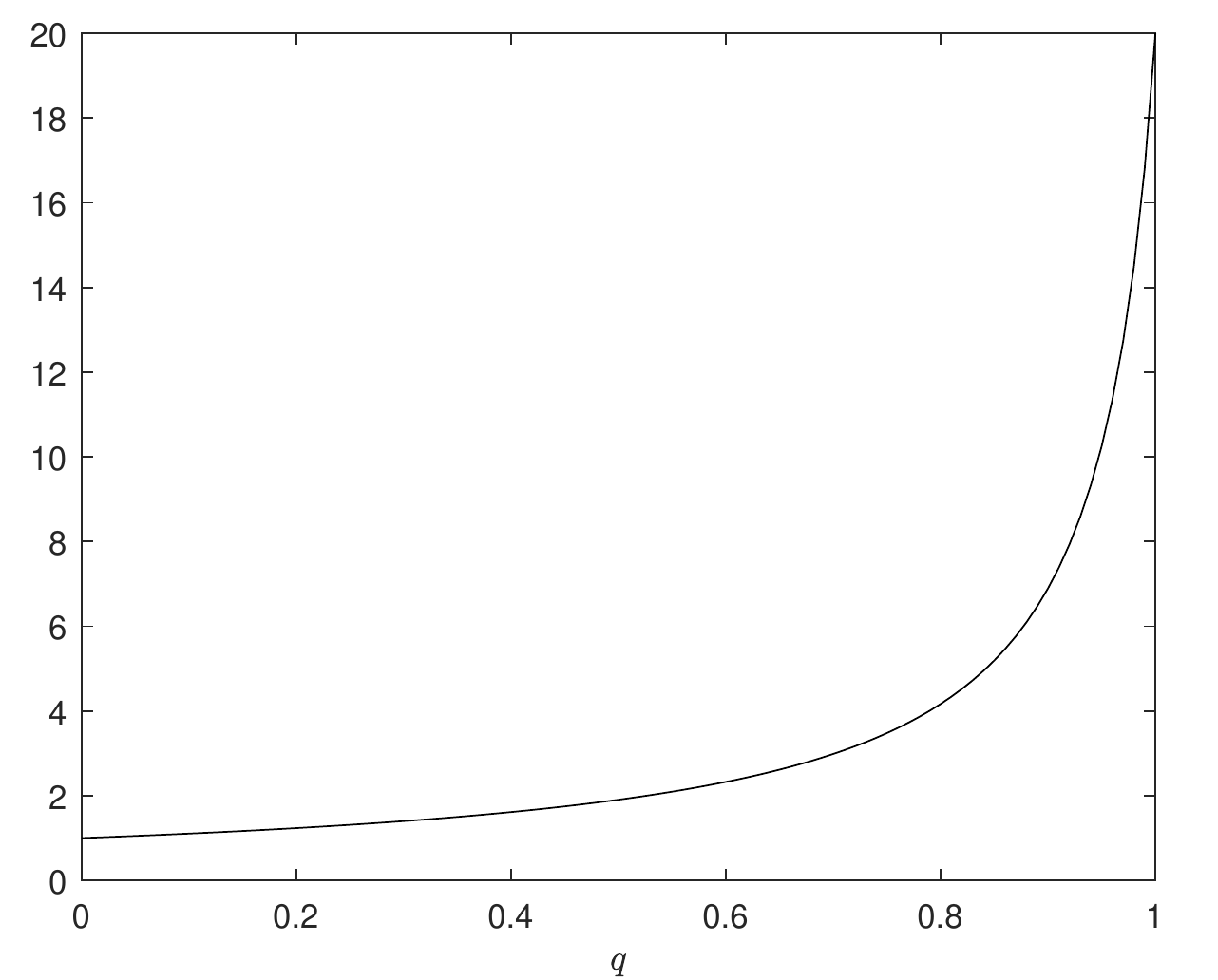}
\caption{Limit of $p(\theta|\widehat T_n>c)/p(\theta)$ as a function of $q$ ($\theta\neq 0$, $\alpha=0.05$)}
\label{fig:probfactor}
\end{figure}

\subsection{Prior with Probability Mass at Zero}
We now consider the case when the prior has probability mass $q$ at zero, with $0<q<1$. Then
\begin{align*}
\Pr\big(\widehat T_n > c\big)&\rightarrow q\alpha+(1-q)\in (\alpha, 1).
\end{align*}
Now, the posterior after significance is,
\begin{align*}
p(0|\widehat T_n > c)&=\frac{\Pr\big(\widehat T_n > c\big| \theta=0\big)}{\Pr\big(\widehat T_n > c\big)}p(0)\rightarrow \left(\frac{\alpha}{q\alpha+(1-q)}\right)q\leq q,\\
\intertext{and}
p(\theta|\widehat T_n > c)&=\frac{\Pr\big(\widehat T_n > c\big| \theta\big)}{\Pr\big(\widehat T_n > c\big)}p(\theta)\rightarrow \left(\frac{1}{q\alpha+(1-q)}\right)p(\theta)\geq p(\theta),
\end{align*}
for $\theta\neq 0$. Now, in contrast to the continuous prior case, significance changes beliefs away from zero in large samples In particular, if we start with a prior that assigns a large probability to $\theta=0$, then significance greatly affects beliefs for values of $\theta$ different from zero. Notice, however, that for moderate values of $q$ the effect of significance on beliefs may be negligible. Figure \ref{fig:probfactor} show the limit of $p(\theta|\widehat T_n>c)/p(\theta)$ as a function of $q$, for $\theta\neq 0$ and $\alpha=0.05$. This limit is close to one for modest values of $q$. In order for significance to at least double the probability of $\theta\neq 0$ we need $q\geq 1/(2(1-\alpha))=0.5263$. Notice that reducing the value of $\alpha$ does not substantially change the value of the limit of $p(\theta|\widehat T_n>c)/p(\theta)$, except for very large values of $q$. For example, with $\alpha = 0.005$ \citep[as advocated in][]{benjamin2017redefine}, for significance to at least double the probability of $\theta\neq 0$ we need $q\geq 1/(2(1-\alpha))= 0.5025$. In fact, regardless of the size of the test, $q$ needs to be bigger than 0.5 in order for significance to double the probability density function of beliefs at non-zero values of $\theta$.

The posterior after non-significance is,
\begin{align*}
p(0|\widehat T_n \leq c)&=\frac{\Pr\big(\widehat T_n \leq c\big| \theta=0\big)}{\Pr\big(\widehat T_n \leq c\big)}p(0)\rightarrow \frac{1-\alpha}{q(1-\alpha)}q=1,\\
\intertext{and for $\theta\neq 0$,}
p(\theta|\widehat T_n \leq c)&=\frac{\Pr\big(\widehat T_n \leq c\big| \theta\big)}{\Pr\big(\widehat T_n \leq c\big)}p(\theta)\rightarrow 0.
\end{align*}
Again, non-significance seems to have a stronger effect on beliefs than significance. 

Some remarks about priors with probability mass at a point null are in order. First, prior beliefs that assign probability mass to point nulls may not be adequate in certain settings. For example, beliefs on the average effect of an anti-poverty intervention may sometimes concentrate probability smoothly around zero, but more rarely in such a way that a large probability mass at zero is a good description of a reasonable prior. Moreover, priors with probability mass at a point null generate a drastic discrepancy, know as  Lindley's paradox, between frequentist and Bayesian testing procedures \citep[see, e.g.,][]{berger1985statistical}. Lindley's paradox arises in settings with a fixed value of $\widehat T_n$ and a large $n$. In those settings, frequentists would reject the null hypothesis when $\widehat T_n>c$. Bayesians, however, would typically find that the posterior probability of the point null far exceeds the posterior probability of the alternative. Lindley's paradox can be explained by the fact that, as $n$ increases, the distribution of the test statistic under the alternative diverges. Therefore, a fixed value of the test statistic as $n$ increases can only be explained by the null hypothesis.
Notice that conditioning on the event $\{\widehat T_n\leq c\}$ (as opposed to conditioning on the value of $\widehat T_n$) is not subject to Lindley's paradox 
and it may be the natural choice to evaluate a testing procedure for which significance depends on the value of $\{\widehat T_n\leq c\}$  only.

\section{Testing an Interval Null}

In view of the lack of informativeness of non-significance in large samples (under a point null), one could instead try to reinterpret significance tests as tests of the implicit null ``$\theta$ is close to zero''.

\begin{figure}
\centering
  \includegraphics[width=4.8in,keepaspectratio=1 ]{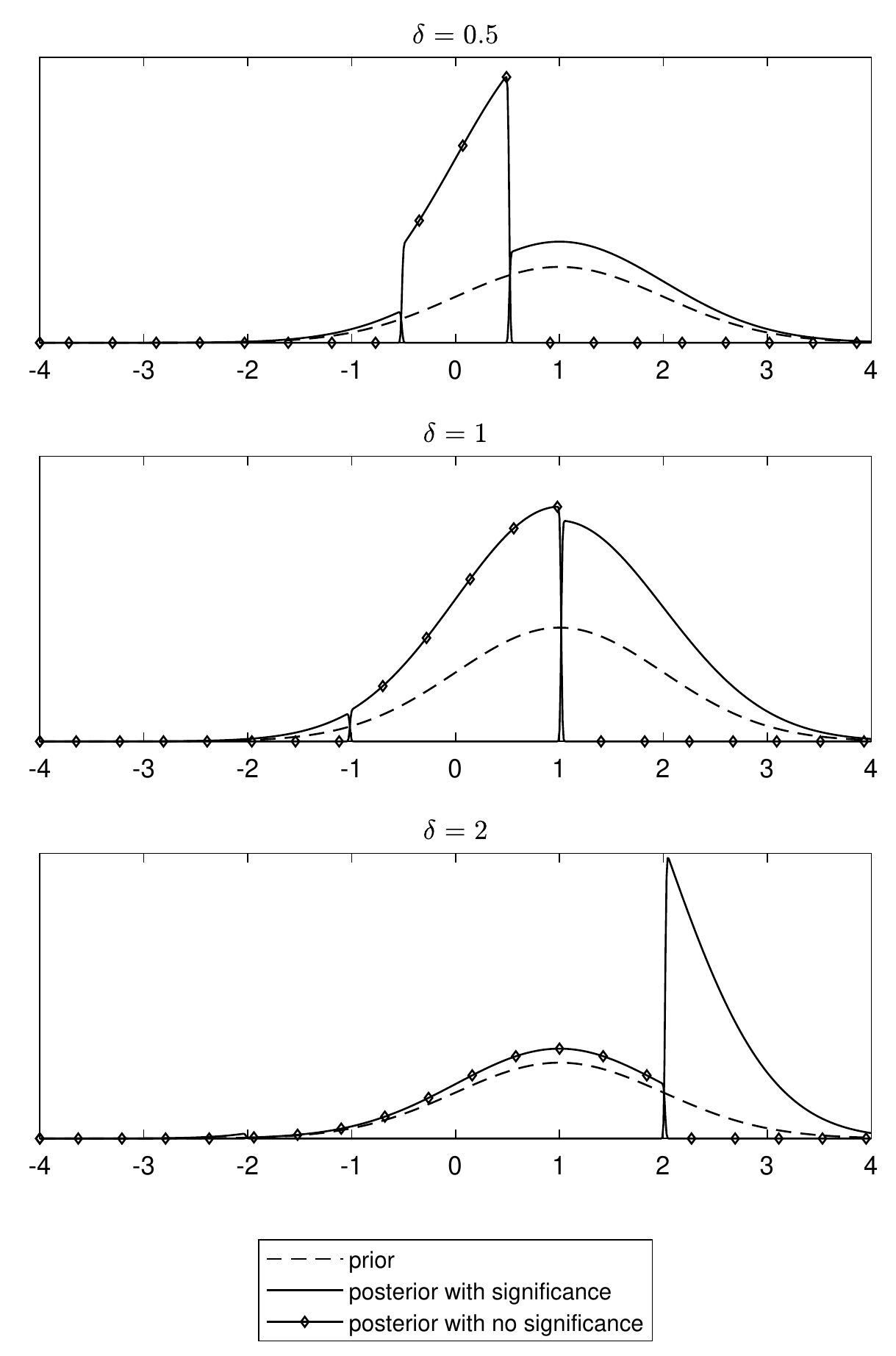}
\caption{Posterior After a Test of the Null $\theta\in [-\delta,\delta]$ ($n=10000$, $\alpha=0.05$)}
\label{figure:interval}
\end{figure}

To accommodate this possibility, we will now concentrate in the problem of testing the null that the parameter $\theta$ is in some interval around zero. Under the null hypothesis, $\theta\in [-\delta,\delta]$, where $\delta$ is some positive number. Under the alternative hypothesis,  $\theta\not\in [-\delta,\delta]$. Consider the Normal model of section \ref{section:normal}. To obtain a test of size $\alpha$ we control the supremum of the probability of Type I error:
\begin{align*}
\Pr(\sqrt{n}|\widehat\theta|> c\,|\,|\theta|=\delta)&=\Phi(\sqrt{n}\delta-c)+\Phi(-\sqrt{n}\delta-c).
\end{align*}
Therefore, we choose $c$ such that $\Phi(\sqrt{n}\delta-c)+\Phi(-\sqrt{n}\delta-c)=\alpha$. While there is no closed-form solution for $c$, its value can be calculated numerically for any given value of $\sqrt n \delta$, and  a very accurate approximation for large $\sqrt n\delta$ is given by
\[
c= \Phi^{-1}(1-\alpha)+\sqrt n\delta.
\]
That is, controlling size in this setting implies that the critical value has to increase with the sample size at a root-$n$ rate. In turn, this implies that the probability of rejection, $\Pr(\sqrt{n}|\widehat\theta|> c|\theta)=\Phi(\sqrt{n}\theta-c)+\Phi(-\sqrt{n}\theta-c)$ converges to one if $\theta\not\in [-\delta,\delta]$, and converges to zero if $\theta\in (-\delta,\delta)$. As a result, the large sample posterior distributions with and without significance are truncated versions of the prior, with the prior truncated at $(-\delta,\delta)$ under significance, and at $(-\infty,-\delta)\cup (\delta,\infty)$ under no significance.  If $\delta$ is large both significance and non-significance are informative. If, however, $\delta$ is small, we go back to the setting where significance carries only local-to-zero information. Figure \ref{figure:interval} reports posterior distributions for $\delta=\{0.5,1,2\}$, $\alpha=0.05$  and $n=10000$.

\section{Conditioning on the sign of the estimated coefficient}
In previous sections we have shown that statistical significance may carry very little information in large samples. As a result, the values of other statistics should be taken into account along with significance when the null is rejected in a significance test. As discussed above, in a Normal (or asymptotically Normal) setting it does not take much to go back to full information (e.g., $P$-value and the sign of $\widehat\theta$). Here we consider the question of whether minimally augmenting the information on significance with the sign of $\widehat\theta$ results in informativeness when the null is rejected. This exercise is motivated by the  possibility that the sign of the estimated coefficient is implicitly taken into account in many discussions of results from significance tests. 

For concreteness, we will concentrate on the case of a positive coefficient estimate, $\widehat\theta>0$. That is, the limited information posterior under significance and positive $\widehat\theta$ conditions on the event $\sqrt n\widehat\theta>c$. The case with negative $\widehat\theta$ is analogous. Using similar calculations as in section \ref{section:introduction}, we obtain:
\[
p\big(\theta\big|\sqrt{n}\widehat\theta>c\big)
=\frac{\displaystyle\frac{1}{\sigma}\phi\Bigg(\displaystyle\frac{\theta-\mu}{\sigma}\Bigg)\Phi(\sqrt{n}\theta-c)}
{\Phi\Bigg(\displaystyle\frac{\sqrt{n}\mu-c}{\sqrt{1+n\sigma^2}}\Bigg)},
\]
and 
\[
p\big(\theta\big|0<\sqrt{n}\widehat\theta\leq c\big)= \frac{\displaystyle\frac{1}{\sigma}\phi\Bigg(\displaystyle\frac{\theta-\mu}{\sigma}\Bigg)\Big(1-\Phi(\sqrt{n}\theta-c)-\Phi(-\sqrt n \theta)\Big)}
{1-\Phi\Bigg(\displaystyle\frac{\sqrt{n}\mu-c}{\sqrt{1+n\sigma^2}}\Bigg)-\Phi\Bigg(\displaystyle\frac{-\sqrt{n}\mu}{\sqrt{1+n\sigma^2}}\Bigg)}.
\]
\begin{figure}
\centering
  \includegraphics[width=4.6in,keepaspectratio=1 ]{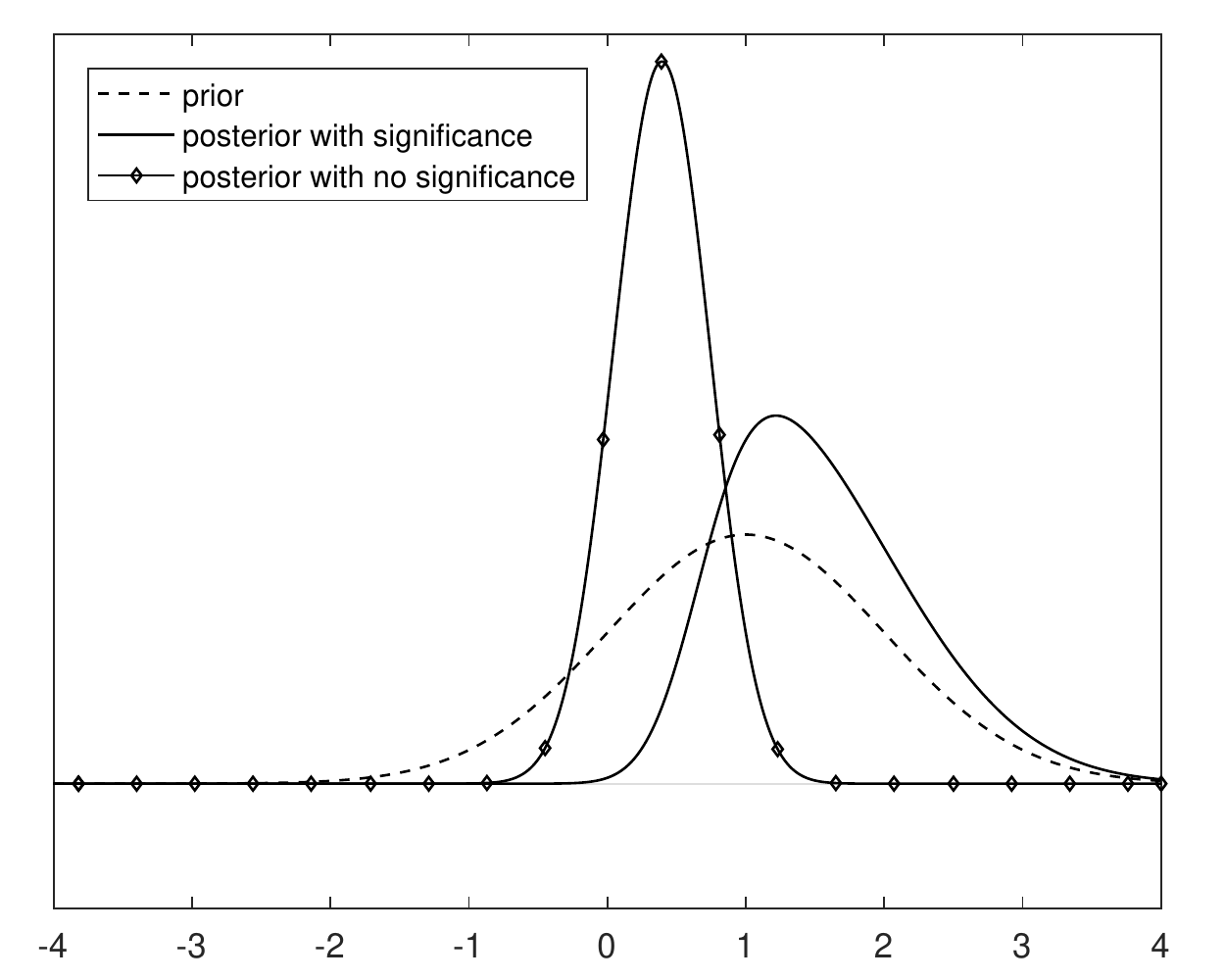}
\caption{Posterior Distributions Conditional of Significance and Coefficient Sign}
\label{fig:posterior_sign}
\end{figure}

Figure \ref{fig:posterior_sign} reproduces the setting of Figure \ref{fig:posterior} but for the case when the posterior is conditional on sign of the estimate in addition to significance. Like in Figure \ref{fig:posterior}, failure to reject carries subtantial information. In fact, both outcomes of the significance test carry additional information, with respect to the setting in Figure \ref{fig:posterior}, which of course is explained by the additional information in the sign of $\widehat\theta$.

Notice that, in this case, under significance, the ratio between the posterior and the prior converges to
\[
\lim\limits_{n\rightarrow \infty}\,\frac{p(\theta|\sqrt n\widehat\theta>c)}{p(\theta)}=\left\{\begin{array}{ll}0&\mbox{ if } \theta<0,\\
\Phi(-c)/\Phi(\mu/\sigma)&\mbox{ if } \theta=0,\\
1/\Phi(\mu/\sigma)&\mbox{ if } \theta>0.\end{array}\right.
\]
Without significance, the ratio between the posterior and the prior converges to
\[
\lim\limits_{n\rightarrow \infty}\,\frac{p(\theta|0<\sqrt n\widehat\theta\leq c)}{p(\theta)}=\left\{\begin{array}{ll}0&\mbox{ if } \theta\neq 0,\\
\infty&\mbox{ if } \theta=0.\end{array}\right.
\]

That is, as $n\rightarrow\infty$ non-significance is highly informative. Under significance, the posterior of $\theta$ converges to the prior truncated at zero. As a result, in this case the informational content of significance depends on the value of $\Pr(\theta>0)=\Phi(\mu/\sigma)$. If this quantity is small, significance with a positive sign is highly informative. Unsurprisingly, when $\mu/\sigma$ is large (that is, in cases where there is little uncertainty about the sign of the parameter of interest), a positive sign of $\widehat\theta$ does not add much to the informational content of the test. Moreover, the limit of $p(\theta|\sqrt n\widehat\theta>c)$ cannot be more than double the value of $p(\theta)$ as long as $\mu$ is non-negative. This is relevant to many instances where there are strong belief about the sign of the estimated coefficients (e.g., the slope of the demand function, or the effect of schooling on wages) and specifications reporting ``wrong'' signs for the coefficients of interest are rarely reported or published.

\section{Conclusions}
Significance testing on a point null is the most extended form of inference. In this article, we have shown that rejection of a point null often carries very little information, while failure to reject is highly informative. This is especially true in empirical contexts where data sets are large and where there are no reasons to put substantial prior probability on a point null. Our results challenge the usual practice of conferring point null rejections a higher level of scientific significance than non-rejections. In consequence, we advocate a visible reporting and discussion of non-significant results in empirical practice \citep[e.g., as in][]{angrist2017maimonides,cantoni2018id}. 
\nocite{*}
\bibliographystyle{chicago}
\bibliography{references}

\end{document}